# A New Interpretation about the Evolution of the Cosmos

**Fang-Pei Chen**
Department of Physics, Dalian University of Technology, Dalian 116024, China.
E-mail: chenfap@dlut.edu.cn

Based on Lorentz and Levi-Civita's conservation laws, it can be shown that the energy of the matter field in the universe might originate from the gravitational field as a result of the latter field's energy decrease and the total entropy increase followed by cosmic expansion. By exploring this possibility and by using some new evidences discovered from recent astronomical observations, we establish an alternative theory of cosmology, which gives a new interpretation about the evolution of the cosmos and a number of new explanations regarding dark energy and dark matter.

**KEY WORDS**: Origin of matter energy; evolution of the cosmos; dark energy; dark matter

## 1. INTRODUCTION

The prevalent theory in cosmologies is the standard big bang cosmology (SBBC), which is based on the theory of general relativity and the cosmological principle that assumes the universe being homogeneous and isotropic in space. A distinguished feature of this theory is that the universe begins to expand from a state of matter field in infinite density called big bang and immediately undergoes a brief period of exponentially fast expansion called inflation. The deductions or predictions of SBBC, such as the cosmic microwave radiation background and the abundance of the helium nuclei in the universe, have been verified by astronomical observations, which help establishing it as the prevalent theory.

In the past two decades numerous new evidences, such as dark energy, dark matter, the strong support for inflation and the total mass-energy density of the universe being close to the critical value, have been accumulated from space surveys. However some new discoveries can not be satisfactorily interpreted by the SBBC theory. For instance, it is difficult or impossible to answer the following questions: Is there really a beginning to the universe? What events led to the onset of inflation [1]? And what are the essence of dark energy and dark matter [2, 3]? These problems make some cosmologists worry that cosmology has "become a victim of its own success" and doubt that the standard model "is less a solid edifice than scaffolding with many gaps resting on uncertain foundations" [1-3]. Other cosmologists think that SBBC is in trouble and it is therefore not premature to give some consideration to alterative cosmologies [4]. These views are well worth considering. In this paper we propose an alternative cosmology in order to give a new interpretation about the evolution of the cosmos.

## 2. OBSAVERTIONAL CONCLUSIONS AND THEORETICAL DEDUCTIONS – FOUNDATIONS TO ESTABLISH AN ALTERNATIVE COSMOLOGY

The following observational conclusions and theoretical deductions will be used as foundations to establish an alternative cosmology; we shall analyze them first.

**2.1 Spatial homogeneity and isotropy**

A high degree of isotropy in cosmological observations, specifically the cosmic microwave radiation background temperature measurements, has been observed [5]. But it is difficult to prove the spatial homogeneity directly in cosmological observations. However astronomical observations tell us that the earth, or the solar system, or our galaxy, or our local group of galaxies, does not occupy any specially favored position in the cosmos, hence we might hypothesize that all positions in the universe are essentially equivalent, *i.e.* the universe is spatial homogeneous.

Using the mathematical property of symmetric space, the Robertson-Walker metric

$$d\tau^2 = -dt^2 + a(t)^2 \{ \frac{dr^2}{1-kr^2} + r^2 d\theta^2 + r^2 \sin^2\theta d\phi^2 \} \qquad (1)$$

for the universe can be deduced immediately [6], where $(r, \theta, \phi)$ are commoving coordinates and *a (t)* is the scale factor. The constant $k$ = -1, 0, 1, is used to indicate the spatial curvature. It has been determined $k$=0 from astronomical observations [3]; but we are not confined to the case $k$=0 only and shall study the general cases at first.

The homogeneous and isotropic assumption also implies that $T_{\mu\nu}$, the energy-momentum tensor of the matter field, should take the form of ideal fluid:

$$T_{\mu\nu} = (\rho + p)u_\mu u_\nu + p g_{\mu\nu}, \qquad (2)$$

where $u_\mu$ is the 4-velocity of matter, $\rho$ is the mean energy density, and $p$ is the mean pressure.

**2.2 Lorentz and Levi-Civita's conservation laws of energy-momentum tensor for gravitational system including matter fields and gravitational fields**

The energy-momentum tensor of the matter field is defined by

$$T_{(M)\mu\nu} \overset{def}{=} \frac{2}{\sqrt{-g}} \frac{\delta(\sqrt{-g} L_M)}{\delta g^{\mu\nu}} \qquad (3)$$

Following the above definition Lorentz and Levi-Civita had defined the energy-momentum tensor for the gravitational field by

$$T_{(G)\mu\nu} \overset{def}{=} \frac{1}{8\pi G\sqrt{-g}} \frac{\delta(\sqrt{-g} L_G)}{\delta g^{\mu\nu}} \qquad (4)$$

The equations of gravitational field can be derived [7] from the requirement that the variation of action integral $I = \int \sqrt{-g}(L_G + 16\pi L_M) d^4x$ should be $\delta I = 0$. Thus they obtained the conservation laws of energy-momentum tensor for gravitational system including matter fields and gravitational fields [8]:

$$T_{(M)\mu\nu}+T_{(G)\mu\nu}=0 \quad \text{and} \quad \frac{\partial}{\partial x^{\mu}}(T_{(M)\mu\nu}+T_{(G)\mu\nu})=0 \qquad (5)$$

we shall call them Lorentz and Levi-Civita's conservation laws.

In the last few years I have thoroughly studied Lorentz and Levi-Civita's conservation laws and found that these conservation laws not only are rational and perfect but also have abundant physical contents [9-12]. A number of new specific properties of gravitational fields or gravitational waves can be deduced and can be tested via experiments or observations [12]. About eighty years ago Einstein did not agree with these conservation laws; the only reason given by him is that these conservation laws "do not exclude the possibility that a material system disappears completely, leaving no trace of its existence."[8], because Einstein believed that the relation expressed by Eq. (5) should make a material system, being $T_{(M)\mu\nu}\neq 0$ in the initial state, to $T_{(M)\mu\nu}\rightarrow 0$ spontaneously. We shall show that this view is not correct. According to statistical mechanics the entropy $S$ of a macroscopic system must obey the Boltzmann's relation $S=k \ln N$, where $N$ is the number of microscopic states. For a macroscopic system, there must be $N \gg 1$ always, thus $S>0$ usually. If a gravitational system (including matter and gravitational field) could disappear completely and spontaneously, then in the disappearing process $N$ will decrease to $N=1$ gradually; at here we look upon the complete disappearance as a special state. Because there is no difference in the meaning between macroscopic and microscopic state for the complete disappearance, so $N=1$. Therefore in the complete disappearance process of this gravitational system its entropy should decrease to $S=0$ from $S>0$; this is contrary to the theorem of entropy increase; hence a gravitational system can not disappear completely and spontaneously.

The energy density of matter field is always positive, so according to Eq. (5) the energy density of gravitational field should be always negative. From Eq. (5) we get $\triangle T_{(M)\mu\nu} = -\triangle T_{(G)\mu\nu}$ immediately, this relation means that for an isolated gravitational system if the energy-momentum of matter field increases, then the energy-momentum of gravitational field should decrease, *i.e.* the energy-momentum of gravitational field might transform into the energy-momentum of matter field. This possibility might occur in reality, since the number of microscopic states both for matter field and gravitational field should all increase in this process so that the entropy of the system increases. It is worth to remember that in the above process the absolute value of gravitational field energy is increasing, thus the number of microscopic states for gravitational field should increase also. This possibility could be used as an important basis for establishing an alternative cosmology.

### 2.3 The cosmological constant $\lambda$, the correction tensor $D_{\mu\nu}$, and the modified Einstein equations for the universe

From astronomical observations we believe that $\lambda\neq 0$ [3, 13], so the term $\lambda g_{\mu\nu}$ should be added in the equations of gravitational field for the universe. On the other hand we have shown in section **2.2** that the energy-momentum of gravitational field might transform into the energy-momentum of matter field; this energy-momentum transformation is equivalent to the creation of matter field's energy-momentum (and the decrease of gravitational field's energy-momentum). The creation of matter field's energy-momentum is a useful concept. This concept had been introduced first in the steady state cosmology [6]; in order to reflect the creation of matter field's energy-momentum, Hoyle had modified the Einstein equations by adding a correction tensor $D_{\mu\nu}$.

Therefore the equations of gravitational field for the universe might be the modified Einstein equations

$$R_{\mu\nu} - \frac{1}{2} g_{\mu\nu} R - \lambda g_{\mu\nu} + D_{\mu\nu} = -8\pi G T_{(M)\mu\nu} \qquad (6)$$

In Eq. (6) the terms $\lambda g_{\mu\nu}$ and $D_{\mu\nu}$ lie on the left. This means that they are similar to $R_{\mu\nu} - \frac{1}{2} g_{\mu\nu} R$ , they are all used to describe the gravitational field; thus $\lambda g_{\mu\nu}$ and $D_{\mu\nu}$ should belong to gravitational field. The gravitational field is different from the matter field: between a gravitational field and a matter field there is only gravitational interaction but between two matter fields there are also other interactions; so it might be possible to distinguish them.

Comparing Eq. (6) with Eq. (5), we get

$$T_{(G)\mu\nu} = \frac{1}{8\pi G} \left( R_{\mu\nu} - \frac{1}{2} g_{\mu\nu} R - \lambda g_{\mu\nu} + D_{\mu\nu} \right) \qquad (7)$$

This equality means $T_{(G)\mu\nu}$ can be divided into three parts:

$$T_{(G)\mu\nu} = \overset{R}{T}_{(G)\mu\nu} + \overset{\lambda}{T}_{(G)\mu\nu} + \overset{D}{T}_{(G)\mu\nu} \qquad (8)$$

where $\overset{R}{T}_{(G)\mu\nu} = \frac{1}{8\pi G} \left( R_{\mu\nu} - \frac{1}{2} g_{\mu\nu} R \right)$ is the part of gravitational field's energy-momentum due to space-time curvature; $\overset{\lambda}{T}_{(G)\mu\nu} = -\frac{\lambda g_{\mu\nu}}{8\pi G}$ is the part of gravitational field's energy-momentum due to cosmological constant; $\overset{D}{T}_{(G)\mu\nu} = \frac{D_{\mu\nu}}{8\pi G}$ is the part of gravitational field's energy-momentum due to the correction tensor $D_{\mu\nu}$. $D_{\mu\nu}$ should be similar to $R_{\mu\nu}$, $g_{\mu\nu}$ and $T_{(M)\mu\nu}$, their non-zero components are only the (0, 0), (1, 1), (2, 2) and (3, 3) component.

The terms $-\lambda g_{\mu\nu}$ and $D_{\mu\nu}$ in Eq. (6) have the property that they look as if they are a part of energy-momentum tensor of the matter fields, for Eq. (6) can be transformed into

$$R_{\mu\nu} - \frac{1}{2} g_{\mu\nu} R = -8\pi G T^{\text{mod}}_{\mu\nu} \tag{9}$$

where $T^{\text{mod}}_{\mu\nu}$ is a modified energy-momentum tensor [6]:

$$T^{\text{mod}}_{\mu\nu} \equiv T_{(M)\mu\nu} - \frac{\lambda}{8\pi G} g_{\mu\nu} + \frac{D_{\mu\nu}}{8\pi G} \tag{10}$$

$T^{\text{mod}}_{\mu\nu}$ could also be written as the perfect-fluid form:

$$T^{\text{mod}}_{\mu\nu} = p^{\text{mod}} g_{\mu\nu} + (p^{\text{mod}} + \rho^{\text{mod}}) U_{\mu} U_{\nu} \tag{11}$$

with a modified density and pressure

$$\rho^{\text{mod}} = \rho_M + \rho_{G\lambda} + \rho_{GD}\, ; \rho_{G\lambda} = \frac{\lambda}{8\pi G},\ \rho_{GD} = \frac{D_{00}}{8\pi G} \tag{12}$$

$$p^{\text{mod}} = p_M + p_{G\lambda} + p_{GD}; \quad p_{G\lambda} = -\frac{\lambda}{8\pi G},\ p_{GD} = \frac{D_{11}}{8\pi G g_{11}} \tag{13}$$

where $\rho_{G\lambda}, \rho_{GD}, \rho_M$ represent respectively the time-time component, *i.e.* the energy density, of $\overset{\lambda}{T}_{(G)\mu\nu}$, $\overset{D}{T}_{(G)\mu\nu}$, $T_{(M)\mu\nu}$ ; $p_{G\lambda}, p_{GD}, p_M$ are the pressure produced by $\overset{\lambda}{T}_{(G)\mu\nu}$, $\overset{D}{T}_{(G)\mu\nu}$, $T_{(M)\mu\nu}$ respectively.

$D_{\mu\nu}$ might be constructed from an unknown field variable, for example, Hoyle had

suggested [6] that $D_{\mu\nu} = C_{;\mu;\nu}$ , C is a scalar field, called C-field. Since we do not yet know the specific properties of the unknown field variable, in this paper we study only the general properties of $D_{\mu\nu}$. The possibility that $D_{\mu\nu}$ might be constructed from an unknown field variable means that the gravitational field variable might not be only the metric tensor $g_{\mu\nu}$, it is therefore possible to include other variable. I believe this possibility and use it as a fundamental hypothesis of the new theory. Of course, whether it is correct or incorrect must be tested by experiments and observations.

2.4 **Some deductions**

. From Eqs. (5-8) we obtain

$$\overset{R}{T}_{(G)\mu\nu} + \overset{\lambda}{T}_{(G)\mu\nu} + \overset{D}{T}_{(G)\mu\nu} + T_{(M)\mu\nu} = 0 \tag{14}$$

$$\frac{\partial}{\partial x^{\mu}}\left(\overset{R}{T}_{(G)\mu\nu} + \overset{\lambda}{T}_{(G)\mu\nu} + \overset{D}{T}_{(G)\mu\nu} + T_{(M)\mu\nu}\right) = 0 \tag{15}$$

then we have

$$\rho_{GR} + \rho_{G\lambda} + \rho_{GD} + \rho_{M} = 0 \tag{16}$$

$$\frac{\partial}{\partial t}\left(\rho_{GR} + \rho_{G\lambda} + \rho_{GD} + \rho_{M}\right) = 0 \tag{17}$$

The value of $\rho_M$ is always positive, *i.e.* $\rho_M \geq 0$; the value of $\rho_{GR}$ is always negative, *i.e.* $\rho_{GR} \leq 0$ ; the value of $\rho_{G\lambda}$ is always positive also, *i.e.* $\rho_{G\lambda} \geq 0$, because $g_{00} = -1$ and if $\lambda > 0$ .We shall explain below that $\rho_{G\lambda}$ might be interpreted as the density of dark energy. As for $\rho_{GD}$, since it will be interpreted as the density of dark matter, we could assume its value is positive, *i.e.* $\rho_{GD} \geq 0$.

. From Eqs. (1), (9), (11-13) we can derive the following two fundamental equations for the scale factor

*a (t)* [6]:

$$\left(\frac{da}{dt}\right)^2 + k = \frac{8\pi G}{3}(\rho_M + \rho_{G\lambda} + \rho_{GD})a^2 \tag{18}$$

$$\frac{d^2a}{dt^2} = -\frac{4\pi G}{3}(\rho_M + \rho_{G\lambda} + \rho_{GD} + 3p_M + 3p_{G\lambda} + 3p_{GD})a \tag{19}$$

The deriving process for these equations might give us a confusing idea that $-\frac{\lambda}{8\pi G}g_{\mu\nu}$ and $\frac{D_{\mu\nu}}{8\pi G}$ are two *true* parts of energy-momentum tensor for the matter fields; however, although $\lambda g_{\mu\nu}$ and $D_{\mu\nu}$ take an active part in the expansion of the universe, yet as it is indicated above that they are not energy-momentum of the matter fields. Eqs. (9) and (10) have only equivalent meaning. Actually the Eqs. (18) and (19) can be also derived from Eqs. (1), (2), (6).

. Besides the above relations, we know that: $(R^{\mu\nu} - \frac{1}{2}g^{\mu\nu}R)_{;\nu}$ =0 and $(\lambda g^{\mu\nu})_{;\nu} = 0$; from Eq. (6) we shall have

$$(D^{\mu\nu} + 8\pi G T^{\mu\nu}_{(M)})_{;\nu} = 0 \tag{20}$$

But there exist the possibilities that $T^{\mu\nu}_{(M);\nu} \neq 0$ and therefore $D^{\mu\nu}_{;\nu} \neq 0$ either; we shall explain that in these cases the energy-momentum might transform between gravitational field and matter field:

Since $T^{\mu\nu}_{(M)} = (\rho_M + p_M)U^\mu U^\nu + p_M g^{\mu\nu}$, we can get [6]

$$T^{\mu\nu}_{(M);\nu} = \frac{\partial p_M}{\partial x^\nu}g^{\mu\nu} + \frac{1}{\sqrt{-g}}\frac{\partial}{\partial x^\nu}[\sqrt{-g}\ (\rho_M + p_M)U^\mu U^\nu]$$

$$+\Gamma^\mu_{\nu\lambda}(\rho_M + p_M)U^\nu U^\lambda \tag{21}$$

For $\mu = t$, then $T^{0\nu}_{(M);\nu} = \frac{\partial \rho_M}{\partial t} + 3\frac{\frac{\partial a}{\partial t}}{a}(\rho_M + p_M)$

$$\approx \frac{\Delta(\rho_M V)}{V\Delta t} + p_M \frac{\Delta V / \Delta t}{V};$$

where V is any volume in the space，$\frac{\Delta(\rho_M V)}{V\Delta t}$ represents the energy change per unit volume per unit time for the matter field, $p_M \frac{\Delta V / \Delta t}{V}$ represents the work done by the matter fluid during the cosmic expansion. If $T^{\mu\nu}_{(M);\nu} = 0$, then $p_M \frac{\Delta V / \Delta t}{V} = -\frac{\Delta(\rho_M V)}{V\Delta t}$; this relation tells us that the whole energy for work done stems from entirely the decrease of matter field energy. If $T^{\mu\nu}_{(M);\nu} \neq 0$, then when $p_M \frac{\Delta V / \Delta t}{V} > -\frac{\Delta(\rho_M V)}{V\Delta t}$, the decrease of matter field energy is less than the work done, so the gravitational field energy must also decrease during the cosmic expansion;

when $p_M \frac{\Delta V / \Delta t}{V} < -\frac{\Delta(\rho_M V)}{V\Delta t}$, the decrease of matter field energy is larger than the work done, some matter field energy must transform to gravitational field energy during the cosmic expansion.

From $(D^{\mu\nu} + 8\pi G T^{\mu\nu}_{(M)})_{;\nu} = 0$ we can derive the relation

$$\frac{\partial}{\partial t}(\rho_{GD} + \rho_M) + 3H(\rho_{GD} + \rho_M + p_{GD} + p_M) = 0 \qquad (22)$$

Where $H(t)=\dfrac{da(t)/dt}{a(t)}$ is the Hubble's constant at time *t*. The equations (16), (17) and (22) can be used to discuss the energy transformations between gravitational field and matter field. From the cosmological principle and using the commoving coordinates we can show that $\rho_{GR}(t)$, $\rho_{G\lambda}(t)$, $\rho_{GD}(t)$, $\rho_M(t)$ and $H(t)$ are all functions of *t* only [6]; if

$\lambda =$ constant, then $\rho_{G\lambda}(t)=$ constant. The initial states of $\rho_{GD}(0)$, $\rho_M(0)$;

$p_{GD}(0)$, $p_M(0)$ at *t=0* and the variation rate $\frac{d}{dt}\rho_{GD}(t)$, $\frac{d}{dt}\rho_M(t)$ have many

possibilities, such as:

**(1)** $\rho_M(0)=0$, $p_M(0)=0$, $\frac{d}{dt}\rho_M(t)>0$, $\frac{d}{dt}p_M(t)>0$; **(2)** $\rho_M(0)>0$,

$\rho_{GD}(0)>0$, $\frac{d}{dt}(\rho_M(t)+\rho_{GD}(t))=0$; **(3)** $\rho_M(0)>0$, $\rho_{GD}(0)>0$,

$\frac{d}{dt}\rho_M(t)>-3H(\rho_M+p_M)$ *etc*. We shall discuss the first two cases as examples.

**Case 1.** $\rho_M(0)=0$, $p_M(0)=0$, $\frac{d}{dt}\rho_M(t)>0$, $\frac{d}{dt}p_M(t)>0$ everywhere

Since $T_{(M)\mu\nu}=p_M g_{\mu\nu}+(\rho_M+p_M)U_\mu U_\nu$, therefore at t=0,

$T_{(M)\mu\nu}(0)=0$ everywhere; *i.e.* the energy-momentum of the matter field is equal to zero in the universe. From Eq. (22) we have

$$\frac{d}{dt}(\rho_M)+3H(\rho_M+p_M)=-[\frac{d}{dt}(\rho_{GD})+3H(\rho_{GD}+p_{GD})].$$

Because $\frac{d}{dt}\rho_M(t) > 0$ and $\frac{d}{dt}p_M(t) > 0$, then $\rho_M(t) > 0$, $p_M(t) > 0$. The

observations of cosmological red shift tell us $H(t) > 0$. Thus the left hand side of the above equation is always larger than zero, which shows that the increase of matter field energy stems from the decrease of gravitational field energy. This means that the energy of matter field might originate from the gravitational field.

**Case 2.** $\rho_M(0) > 0$, $\rho_{GD}(0) > 0$, $\frac{d}{dt}(\rho_M(t)+\rho_{GD}(t))=0$ everywhere

In this case it is evident that $\rho_M(t) + \rho_{GD}(t)$ = constant >0, Eq. (22) tells us

$p_{GD}(t) + p_M(t)$ = constant <0. Besides, if $\lambda>0$, then $p_{G\lambda} = -\frac{\lambda}{8\pi G}$ <0 also, the universe shall expand with negative pressure. Astronomical observations tell us that the space might be flat *i.e.* $k = 0$, from Eq. (18), $\frac{da}{dt}\Big/ a = H$ = constant for all t, therefore

$a(t)=a(t_0)\exp\{H(t-t_0)\}$ [6]. This result means the universe is inflating.

## 3. A NEW THEORY OF COSMOLOGY

The results of discussions in the section 2 are sufficient to establish a new theory of cosmology. The chief contents of the new theory given by this paper are:

(1). It accepts the cosmological principle, *i.e.* the universe is spatial homogeneous and isotropic, so the universe has the Robertson-Walker metric

$$d\tau^2 = -dt^2 + a(t)^2\left\{\frac{dr^2}{1-kr^2} + r^2 d\theta^2 + r^2\sin^2\theta d\phi^2\right\} \qquad (1)$$

Astronomical observations indicate the universe is spatial flat, *i.e.* $k=0$; then Eq. (1) become

$$d\tau^2 = -dt^2 + a(t)^2\{dr^2 + r^2 d\theta^2 + r^2\sin^2\theta d\phi^2\} \qquad (1\text{-a})$$

(2). It adopts the modified Einstein equations

$$R_{\mu\nu}-\frac{1}{2}g_{\mu\nu}R-\lambda g_{\mu\nu}+D_{\mu\nu}=-8\pi GT_{(M)\mu\nu} \qquad (6)$$

as the equations of gravitational field for the universe. These equations and the conclusions deduced from them apply to the whole cosmos, for the entire cosmos $\rho_{G\lambda},\rho_{GD},\rho_M$ are all less than the critical density [6] $\rho_C=\frac{3[H(t_0)]^2}{8\pi G}$ = $1.9h^2\times10^{-29}\,g/cm^3$ . But for a macroscopic gravitational system

$\rho_M$ >> $\rho_C$ and $\rho_{G\lambda},\rho_{GD}$ still less than $\rho_C$ (but from Eq. (16) $\rho_{GR}\approx-\rho_M$ ),

therefore Eq. (6) degenerates to $R_{\mu\nu}-\frac{1}{2}g_{\mu\nu}R=-8\pi GT_{(M)\mu\nu}$ .

(3). It uses the Lorentz and Levi-Civita's conservation laws as one of its theoretical foundations. It means that the energy-momentum of matter field might create from gravitational field.

The equations $T_{(M)\mu\nu}+T_{(G)\mu\nu}=0$ tell us, when $T_{(M)\mu\nu}=0$, $T_{(G)\mu\nu}=0$. In this state, $\rho_M$ = 0, $p_M$ = 0, this is the lowest state of energy-momentum for the matter field in the universe. It must indicate that this state does not equal to the other 'lowest' energy state of pure matter field, *i.e.* the so called 'vacuum' state, of quantum matter field; at the 'vacuum' state, $\rho_M>0$. It must indicate also that the energy creation of a matter field does not mean the matter field creation, thus if at $t=0$, $\rho_M(0)$ = 0, at $t>0$, $\rho_M(t)>0$, it means only that the state of matter field does change from the lowest energy state to a higher energy state, the concept of matter field creation is not necessary.

Why $\rho_M$ = 0, $p_M$ = 0 is the lowest state of energy-momentum for the matter field in the universe? How is the energy-momentum transformed from the gravitational field into the matter field? These problems relate with the quantum theory of gravitational field. On account of a complete and consistent quantum theory of gravitational field has not been constructed yet till now, so we can not reply fully these problems at once now.

**(4).** It hypothesizes that $\rho_{G\lambda}$ is the density of dark energy and $\rho_{GD}$ is the density of dark matter.

Astronomical measurements suggest that the expansion of the universe is accelerating; using the hypothesis (4) we can easily explain the accelerating expansion of the universe.

Because $p_{G\lambda} = -\frac{\lambda}{8\pi G} < 0$, therefore, from Eq. (19) if

$$(\rho_M + \rho_{G\lambda} + \rho_{GD} + 3p_M + 3p_{GD}) < -3p_{G\lambda} \text{ ; then } \frac{d^2 a}{dt^2} > 0 .$$

From Eq. (18) and $k = 0$ we can get the relation [13]:

$$\Omega_M + \Omega_{G\lambda} + \Omega_{GD} = 1 \tag{23}$$

where $\Omega_M = \frac{\rho_M}{\rho_C}, \Omega_{G\lambda} = \frac{\rho_{G\lambda}}{\rho_C}, \Omega_{GD} = \frac{\rho_{GD}}{\rho_C}$; $\rho_C = \frac{3[H(t_0)]^2}{8\pi G}$ is the critical density [6]; $t_0$ is the time of the present moment. The conclusions from CMB data tell us that [3] the Universe has 73% dark energy, 23% dark matter and 4% ordinary (baryonic) matter. According the above hypothesis we would have: $\rho_{G\lambda}/\rho_{\mathrm{mod}} = 73\%$ , $\rho_M/\rho_{\mathrm{mod}} > 4\%$ , $\rho_{GD}/\rho_{\mathrm{mod}} < 23\%$ and $\rho_M/\rho_{\mathrm{mod}} + \rho_{GD}/\rho_{\mathrm{mod}} = 27\%$ ; because some parts of the 'dark matter' might be material matter [3], such as the neutrino, a weakly interacting massive particle (WIMP) and the massive compact halo objects (MACHOs, including low-luminosity stars and black holes). In the new theory the other part of the dark matter should be the field of $D_{\mu\nu}$ with energy density $\rho_{GD}$. We have considered that the field $D_{\mu\nu}$ should be a part of gravitational field in essence, so the properties of $\rho_{GD}$ would be different from $\rho_M$. Their differences might be tested by experiments and observations in the future.

It is reasonable to interpret $\rho_{G\lambda}$ as the density of dark energy. In addition to the explanation of the accelerating expansion of the universe as shown above, the term $-\lambda g_{\mu\nu}$ can also be used to explain the universe's inflation. The inflation stage is necessary for the standard big bang cosmology but is not necessary for the new theory of cosmology established in this paper.

$\rho_{G\lambda}$ is a part of gravitational field's energy density and belongs to $-\frac{\lambda}{8\pi G}g_{\mu\nu}$, the vacuum energy density is a part of matter field's energy density and belongs to $T_{(M)\mu\nu}$, thus they might be different in essence. So it appears that there is no definitive relation between $\rho_{G\lambda}$ and the vacuum energy density. Such a relationship was assumed to exist [13], but such an assumption all led to some form of difficulties and complications. In the new theory of cosmology it is not necessary to establish the relation between $\rho_{G\lambda}$ and the vacuum energy density.

The SBBC has a starting state called big bang and assumes that the total energy of matter fields (including the inflation field) has existed from the big bang; moreover, this theory does not study the origin of the matter field's energy. The new theory of cosmology established in this paper has no big bang, it is without a beginning and without an end; the space expands continuously. The view of no beginning means that the state *t=0,* $\rho_M = 0$ does not exist. Why isn't there a beginning state *t=0,* $\rho_M = 0$? This is due to the quantum fluctuations, at any time there must always be energy-momentum transformation between gravitational field and matter field, so the beginning state *t=0,* $\rho_M = 0$ is not possible to appear.

The steady state cosmology introduced firstly the concept regarding the creation of matter field's energy and modified firstly the Einstein equations by adding a correction tensor $D_{\mu\nu}$; the quasi- steady state cosmology, which is a revised theory of steady state cosmology, adopted firstly the modified Einstein equations, *i.e.* Eq. (6), with the term $\lambda g_{\mu\nu}$. However our new theory of cosmology is different in principle from the above two cosmologies: the new theory uses the Lorentz and Levi-Civita's conservation laws as one of its theoretical foundations, but the other two do not; the steady state cosmology affirms that there are $\frac{d}{dt}(\rho_M(t)+\rho_{GD}(t))=0$ everywhere and $H=$ constant for all t (see **2.4 case 2**), but the new theory does not; the quasi-steady state cosmology assumes that

$$D_{\mu\nu}=-\frac{2}{3}(C_\mu C_\nu-\frac{1}{4}g_{\mu\nu}C_\sigma C^\sigma) \qquad (24)$$

$C_\mu$ is a vector field; but the new theory keeps $D_{\mu\nu}$ in general form, so that Eq. (24) is only its special case, and $D_{\mu\nu}$ could suit many cases.

It is well known that the observations of the cosmic microwave radiation background and the observed abundances of light nuclei in the universe caused many cosmologists to favor the standard cosmology (SBBC) over the steady state cosmology; since SBBC can explain these two events well, but the steady state cosmology does not. In SBBC the observed abundances of light nuclei in the universe are explained as the result of nucleon-synthesis taking place in a very hot dense stage after the big bang. There are other explanations about the observed abundances of light nuclei in the universe. Some cosmologists in the 1950's had studied the possibilities that the light nuclei in the universe are formed from hydrogen in the interiors of stars [6]. But the cosmic abundance of helium is too large to be easily explained in terms of nucleon-synthesis in the interiors of stars at $10^{10}$ years estimated by SBBC. However the new theory of cosmology is without a beginning state, the helium nuclei in the universe might have synthesized for a very long time; therefore the above problem does not exist. In SBBC the cosmic microwave radiation background is interpreted as the relic of the early hot era. There are other explanations also about the cosmic microwave radiation background; even for the steady state cosmology, Weinberg said: " it is not out of the question for a microwave background to be created along with the baryons in a steady state model", although the Plank distribution law is possible but quite artificial. The quasi-steady state cosmology maintains [4] that the microwave background is the thermalized relic starlight left by stars that have burnt during the ancient times and considers that the iron whiskers can act as efficient thermalizers of starlight. I shall adopt the above viewpoints of the quasi- steady state cosmology, although there are some problems in their calculations; these problems might be corrected by changing the initial states of $\rho_{GD}(0)$, $\rho_M(0)$; $p_{GD}(0)$, $p_M(0)$ at *t=0* and the variation rate $\frac{d}{dt}\rho_{GD}(t)$, $\frac{d}{dt}\rho_M(t)$ which have many possibilities as we have seen in **2.4**.

Whether a theory is correct or not, it must be tested by experiments and observations. We outline a few tests that could either confirm or disprove the new theory of cosmology in the

following:

**1). Testing the Lorentz and Levi-Civita's conservation laws**

Various concrete experiments and observations using the specific properties of gravitational waves to test the Lorentz and Levi-Civita's conservation laws were enumerated in Ref. [12]. These conservation laws are the foundation of the new theory; their correctness means that the energy-momentum of the matter field might create from gravitational field. So that to confirm these conservation laws is to confirm indirectly the new theory of cosmology and to disprove SBBC, since SBBC does not permit the creation of matter field's energy-momentum. The quasi-steady state cosmology does not adopt the Lorentz and Levi-Civita's conservation laws but permit the creation of matter field's energy-momentum; therefore to confirm these conservation laws is neither to confirm nor to disprove the quasi-steady state cosmology.

**2). Probing into the essence of dark matter**

We have explained above that in the new theory of cosmology some parts of the 'dark matter' might be material matter, the other part of the dark matter should be the field of $D_{\mu\nu}$ with energy density $\rho_{GD}$ and $\rho_M / \rho_{\mathrm{mod}} > 4\%$ , $\rho_{GD} / \rho_{\mathrm{mod}} < 23\%$ and $\rho_M / \rho_{\mathrm{mod}} + \rho_{GD} / \rho_{\mathrm{mod}} = 27\%$ . $D_{\mu\nu}$ is a part of gravitational field. The gravitational field is different from the matter field, $\rho_{GD}$ *and* $\rho_M$ can interact with gravitational force but can not interact with other forces, so it might be possible to distinguish them. But SBBC and quasi-steady state cosmology all consider the dark matter as some kinds of matter field [4], hence to confirm the above viewpoints is to confirm the new theory of cosmology and to disprove SBBC and the quasi-steady state cosmology.

**3). Probing into the essence of dark energy**

We have indicated above also that $-\lambda g_{\mu\nu}$ is a part of gravitational field, $\rho_{G\lambda}$ *and* $\rho_M$ can interact with gravitational force but can not interact with other forces, so it might be possible to distinguish them as to distinguish $\rho_{GD}$ *and* $\rho_M$ .

**4). Finding very old stars**

The new theory of cosmology is without a beginning and without an end; therefore very old stars must exist. To find very old stars is necessary to confirm the new theory of cosmology and to disprove SBBC. But this test in itself does not distinguish the new theory from the quasi-steady state cosmology or the steady state cosmology as they are all without a beginning state.

## 4. CONCLUDING REMARKS

As it has been explained in the introduction section of this paper, the new evidences of observations have brought out some crucial weaknesses of SBBC. It is necessary to introduce new concepts and new laws; the main objective of this paper is to show such necessity and to derive a new alternative theory of cosmology. The current work is only preliminary and it is hoped that this work may generate further interests and studies in establishing a better alternative theory of cosmology.

**REFERENCES**

1. Gratton, S. and Steinhardt, P. (2003). *Nature* **423,** 817.
2. Carroll, S. (2003). *Nature* **422,** 26.
3. Fukugita, M. (2003). *Nature* **422,** 489.
4. Narlikar, J. V. (1999). *Pramana-J.Phys.***53,** 1093.
5. Ellis, G. F. R. (1999). *Gen. Rel. Grav.***32,** 1135.
6. Weinberg, S. (1972). *Gravitation and Cosmology.* (Wiley. New York).
7. Carmeli, M. (1982). Classical Fields: General Relativity and Gauge Theory. (Wiley. New York).
8. Cattani, C.and De Maria, M. (1993). "Conservation laws and gravitational waves in general relativity". In: *The Attraction of Gravitation.* Edited by Earman, J., Janssen, M. and Norton, J. D. (Birkhauser. Boston).
9. Chen, F. P. (1997). *Journal of Dalian University of Technology.* **37,** 33. (In Chinese)
10. Chen, F. P. (1998). *Ziran Zazhi (Journal of Nature)* .**20,** 178. (In Chinese)
11. Chen, F. P. (2000). *Journal of Hebei Normal University.* **24,**326. (In Chinese)
12. Chen, F.P. (2002). *Spacetime & Substance.* **3,** 161.
13. Peebles, P. J. E. and Ratra, B. (2003). *Rev. Mod. Phys.* **75,** 559.